# Intermolecular orientations in liquid acetonitrile: new insights based on diffraction measurements and all-atom simulations

Szilvia Pothoczki and László Pusztai
*Institute for Solid State Physics and Optics, Wigner Research Centre for Physics, Hungarian Academy of Sciences, Konkoly-Thege M. út 29-33, 1121, Budapest, Hungary*

**Abstract**

Intermolecular correlations in liquid acetonitrile ($CH_3CN$) have been revisited by calculating orientational correlation functions. In the present approach, hydrogen atoms are included, so that a concept applicable for molecules of (nearly) tetrahedral shape can be exploited. In this way molecular arrangements are elucidated not only for closest neighbours but also extending well beyond the first coordination sphere. Thus a complementary viewpoint is provided to the more popular dipole-dipole correlations. Our calculations are based on large structural models that were obtained by applying diffraction data and partial radial distribution functions from potential-based (all-atom) molecular dynamics simulation simultaneously, within the framework of the Reverse Monte Carlo method.

## 1. Introduction

Liquid acetonitrile has attracted a continuous interest[1-24] over the past nearly 40 years, due to its physical properties (high dipole moment, high dielectric constant, miscibility with protic solvents) that allow for a wide range of applications[1]. Its structure has been investigated both by X-ray[2-4] and neutron[5-6] diffraction methods. In these early studies even the (intermolecular) partial radial distribution functions were not determined. Thus it became clear that for detailed analyses of molecular-level correlations, the application of computational methods, such as molecular dynamics (MD)[7-15], Monte Carlo (MC)[16-20] and reverse Monte Carlo (RMC) simulation[18], and/or various theoretical calculations based on statistical mechanics[21-24], would be necessary. For determining mutual orientations of the molecules in the liquid, structural models containing thousands of molecules would be essential.

Unfortunately, it is difficult to assess statements and findings arising from these calculations because the majority of the works mentioned above[7,9-13,15-17,19-24] have not considered any comparison with any diffraction measurements[2-6] (as direct information of the structure). As a reminder, we wish to point out here that just one type of diffraction data, either X-ray or neutron, cannot provide the appropriate information necessary for determining even all the two-body intermolecular correlations[14,18]: X-ray diffraction is mainly sensitive to carbon-carbon and nitrogen-nitrogen correlations, whereas neutron diffraction is most sensitive to pair correlations involving hydrogen atoms.

Accordingly, *our first aim is to generate large structural models that are consistent with both neutron- and X-ray diffraction data*. For this purpose we apply the Reverse Monte Carlo (RMC) technique[25]. We note here that one RMC based study can be found in the literature[18], from the early years of the method, using only X-ray diffraction data and molecules of only three sites, i.e., without explicitly including H atoms. Similarly, most of the previous structural models (except for Refs. 7, 8, 11-13) do count hydrogen atoms separately but just a 'united atom type' methyl group. One of the novelties of the present study is that *our approach, while making use of also neutron diffraction data, considers hydrogen atoms of the methyl group explicitly,* with the aim of gaining information about intermolecular correlations, including orientations, between realistic molecules. This allows acetonitrile molecules to be taken as elongated tetrahedra, with the nitrogen and the three hydrogen atoms as the four corners. Distance-dependent orientational correlation functions can be calculated to describe mutual orientations of these distorted tetrahedra, similarly to the case of liquid chloroform[26]. This is a possible way of revealing orientational correlations between molecules that are most frequently handled as linear bodies (possibly with dipolar vectors defined along their axes).

A more traditional way of determining mutual arrangements of dipolar molecules is to calculate correlation functions using the angle confined between two dipole vectors[8,16-20]. Here we also provide distance dependent dipole-dipole correlation functions, for comparison with earlier findings[8,16-20]. Furthermore, we aim to introduce two additional characteristic angles[27,28], in order to complement the standard description of dipole-dipole correlations so that, for instance, within antiparallel arrangements 'head-to-head' and 'tail-to-tail' type orientations may become distinguishable. These orientations are hardly identifiable without this extra.

In addition, one further question emerges: can we say anything about *mutual orientations beyond the first coordination shell*? The present study reports an attempt to address this issue.

A general difficulty concerning multi-component systems (in our case, at least the three constituents, N, C and H, need to be taken as 'components') is encountered here: due to the 'all-atom' approach the number of available independent diffraction data sets is lower than the number of partial radial distribution functions. To handle this kind of a lack of information, partial radial distribution functions from MD simulations have been utilized as input data, together with the experimentally determined total scattering structure factors (TSSF) of neutron and X-ray diffraction. Perfect agreement (within experimental uncertainties) with diffraction data was required from the RMC calculations. The expectation against partial radial distribution functions (PRDF) from MD was to see how well the potential-based PRDF-s can be approached while fitting experimental data perfectly. A combined RMC+MD scheme, suggested some years ago[29] and described in more detail recently[30], is, apart from potentially improving the quality of RMC structures, also a possible tool for a detailed validation of interaction potentials used in MD.

The initial configuration of our RMC calculations has been constructed by means of 'all-atom' Molecular Dynamics simulation using a 6-site potential model for acetonitrile. Every molecular site in the MD corresponded to an atom in RMC. During the RMC simulation all atoms were treated separately from each other. Particle configurations from RMC and also from MD simulations could later be analysed in detail; here, correlation functions mentioned above have been computed (see Section 2.3 for details).

It was found important to provide comparison with the most relevant findings of earlier studies,[8,13,14,16,18] especially of the work of Böhm et al.[8], as this latter work formulated objectives fairly to close our intentions.

The paper is organized as follows. In Sec II, we briefly introduce our method for the preparation of structural models exploited in this study, accompanied by a detailed description of the three different orientational correlation functions that our results are based on. The evaluation of the models can be found in Sec. III. Results regarding intermolecular correlations are summarized in Section IV, and conclusions are presented in Sec. V.

## 2. Computational details

We obtained our final ('ready-for-analyses') model in two steps: (1) first a potential based Molecular Dynamics simulation has been performed; (2) then the final configuration of the MD simulation became the initial configuration of reverse Monte Carlo (RMC) modelling, by which we wished to refine the MD structure.

Partial radial distribution functions coming from the initial MD simulation were used as input data for RMC, just as measured total scattering structure factors from diffraction experiments. Due to chemical considerations, the two carbon atoms within the $CH_3CN$ molecule were distinguished and therefore in practice—with the N and H atoms—we had a four-component system. Consequently, the number of partial contributions (10) was much larger than the number of available diffraction data sets (2), leading to the information deficiency mentioned earlier.

## 2.1 Molecular dynamics simulation

Molecular dynamics simulations have been performed in the NVT ensemble using the GROMACS 4.0 program package[31] at T = 293 K. The temperature was controlled by the Berendsen thermostat[32], with the temperature coupling time constant $\tau$ set to 0.1 ps. The initial configuration contained 2000 molecules (12000 atoms), with randomly placed molecular centres, in a cubic simulation box with periodic boundary conditions. The edge length of the simulation box was 55.837 Å, corresponding to the experimental density (0.786 g/cm$^3$). The OPLS all-atom force field[33] was selected for representing interactions between molecules. σ parameters of the Lennard-Jones (LJ) potential were 3.3 Å ($C_{methyl}$ and C), 2.5 Å (H) and 3.2 Å (N), whereas LJ ε-s were 0.276144 kJ/mol ($C_{methyl}$ and C), 0.06276 kJ/mol (H) and 0.711280 kJ/mol (N). The partial charges were distributed, so that they bring about the correct dipole moment (3.92D): -0.08 e ($C_{methyl}$); 0.46 e (C); 0.06 e (H); -0.56 e (N) (i.e., the units are of the elemental charge). The calculation of the non-bonded interactions was optimized by a grid-based neighbour list algorithm (the lists were updated in every 10 steps). Both electrostatic and van der Waals interactions were truncated at 0.9 nm, and the particle mesh Ewald method[34] was employed for the long-range electrostatic interactions. Bond lengths and angles were kept flexible using the LINCS algorithm[35], allowing for an integration time step of 2 fs. Initial bond lengths were set to 1.47 Å ($C_{methyl}$-C), 1.157 Å (C-N), and 1.09 Å ($C_{methyl}$-H), accompanied by initial bond angles of 180° ($C_{methyl}$-C-N), and 107.8° (H-$C_{methyl}$-H).

The total simulation time was 2000 ps. A steepest-descent gradient method was applied prior to the simulations for energy minimization and to avoid atomic overlaps in the system. The total energy reached its equilibrium value within 100 ps. Data from the last 1500 ps were used for further analyses, e.g., for calculating partial radial distribution functions, and comparison with scattering data. For the latter, the *g_rdf* software of the GROMACS package was modified; details of the calculation can be found in Ref 36.

## 2.2 Reverse Monte Carlo modelling

The reverse Monte Carlo method[25,37-39] is a way to generate large structural models that are consistent with experimental data within their errors. In this study we applied partial radial distribution functions arising from MD simulation (see previous section) simultaneously with total scattering structure factors from X-ray and neutron diffraction experiment during the fitting procedure. Although earlier X-ray diffraction data can be found in the literature[2-4], we took results from a new experiment[40] that has been carried out at the SPring-8 synchrotron radiation facility (Japan), using the single-detector diffractometer setup of the BL04B2 (high-energy X-ray diffraction) beamline[41]. Neutron diffraction measurements were taken from Ref. 5.

Our RMC simulations were started from particle configurations resulting from the preceding MD simulation (see previous section). In this way, RMC calculation may be considered as a "refinement" of the MD results. The basics of RMC modelling can be found in, e.g., Refs.37-39, therefore only the relevant details are provided here. The atomic number density (0.06893 atom/Å$^3$) and the simulation box lengths (55.837 Å) were identical in MD and RMC simulations. Molecules have been kept together by means of 'fixed neighbours constraints' (fnc)[38], which keep atoms within a molecule within pre-specified minimum and maximum distances; in our case, specifically: 1.44-1.48 Å ($C_{methyl}$-C), 0.987-1.187 Å ($C_{methyl}$-H), 1.99-2.19 Å (C-H), 1.665-1.875 Å (H-H), 2.57-2.69 Å ($C_{methyl}$-N ), 1.15-1.19 Å (C-N), 3.1-3.22 Å (H-N). In addition, the $C_{methyl}$-C-N bond angle has been required to be 180° with a small tolerance.

The present RMC calculations were run on the basis of atomic movements: this is why the tolerances for the intramolecular (FNC) distances had to be set as relatively wide – otherwise hardly any attempted moves could be accepted. Still, the movement of entire molecules was restricted; this is why the molecular dynamics algorithm had been applied before RMC, so that the system was allowed to explore the 'configurational space' prior to the final RMC refinement. In this particular case, one such 'MD—RMC' cycle proved to be sufficient as already the MD simulation provided a food agreement with experimental data (cf. Figure 2).

To prevent overlaps of the atoms, the following closest approach values were enforced: 3.0

Å ($C_{methyl}$-$C_{methyl}$), 3.0 Å ($C_{methyl}$-C), 2.5 Å ($C_{methyl}$-H), 2.8 Å ($C_{methyl}$-N), 2.9 Å (C-C), 2.4 Å (C-H), 2.7 Å (C-N), 2.0 Å (H-H), 2.2 Å (H-N), 3.0 Å (N-N).

The essence of the present study is the analyses based on four different kinds of orientational correlation functions (see the next section for details); all these characteristics have been calculated directly from particle coordinates.

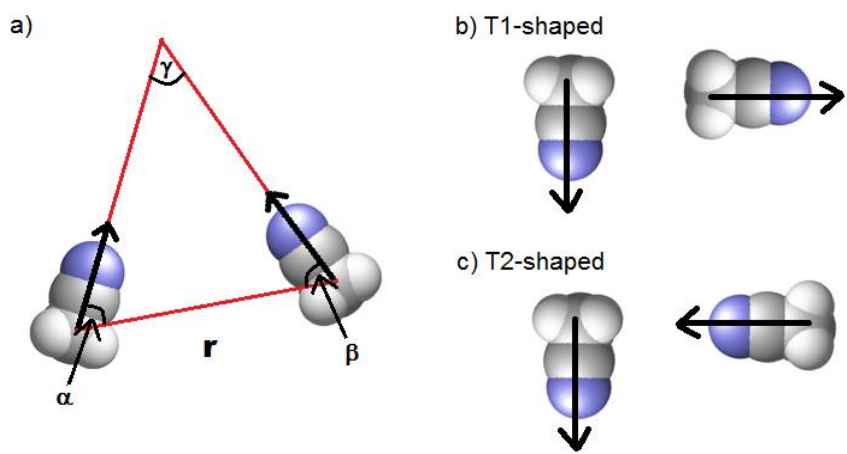

Figure 1. a) Schematic representations of characteristic angles describing special orientations for acetonitrile. b) "T1-shaped" (cosα=0, cosβ=-1, cosγ=0) arrangement. c) "T2-shaped" (cosα=1, cosβ=0, cosγ=0) arrangement.

**2.3 Correlation functions for characterizing mutual orientations of the molecules**

To characterize the mutual orientations of the molecules the following correlation functions have been used:

(1) *Dipole–dipole correlation functions*. The cosines of angles confined between two dipole vectors have been calculated as a function of the distance between two molecules. Note also that in what follows, it is *not* the average, but always the individual (cosine of the) angle that is taken into account.

(2) S*pecific dipole-dipole correlation functions.* In addition to the simple dipole–dipole angle, introduced previously, angles confined by the molecular axes and the line connecting molecular centres have also been calculated. With the help of these three angles the following eight orientations have been monitored (similarly to recent studies[26,27]): "parallel" (cosα=0, cosβ=0, cosγ=1), "antiparallel" (cosα=0, cosβ=0, cosγ=-1), "T1-shaped" (cosα=0, cosβ=-1, cosγ=0), "T2-shaped" (cosα=1, cosβ=0, cosγ=0) "head-to-head" (cosα=1, cosβ=1, cosγ=-1), "head-to-tail" (cosα=1, cosβ=-1, cosγ=1), "tail-to-tail" (cosα=-1, cosβ=-1, cosγ=-1) and "crossed-shaped" (cosα=0, cosβ=0, cosγ=0). Figure 1 provides a definition of the angles in question and shows the T1 and T2 orientations explicitly.

(3) *Orientational correlation functions for tetrahedral molecules*[42]. The acetonitrile molecule can be considered as one with the shape an elongated tetrahedron (remember that the N-C-C backbone is on a straight line). From this point on the original construction of these correlation functions[42] is applied in the following way: two parallel planes are constructed that contain the centres (here, methyl carbon atoms) of the two molecules in question and that are perpendicular to the line joining the molecules. Every molecular pair may be classified, as a function of the distance between centres of the molecules, by the number of ligands between the planes into one of the following groups: corner-to-corner (1:1), corner-to-edge (1:2), edge-to-edge (2:2), corner-to-face (1:3), edge-to-face (2:3) and face-to-face (3:3) orientations.

(4) *Subgroups* of the original 'tetrahedral' orientational correlation functions. By elaborating the original idea, the two types of ligands that constitute the corners of tetrahedra (hydrogens and nitrogen) may be distinguished. In this way, 21 subgroups result for the $CH_3CN$ liquid (similarly to the case of chloroform)[26]. The complete list of subgroups is as follows: 1:1 {(H-H), (H-N), (N-N)}; 1:2 {(H-H,H), (H-H,N), (N-H,H), (N-H,N)}; 2:2 {(H,H-H,H), (H,N-H,H), (H,N-H,N)}; 1:3 {(H-H,H,H), (N-H,H,H), (H-H,H,N), (N-H,H,N)}; 2:3 {(H,H-H,H,H), (H,N-H,H,H), (H,H-H,H,N), (H,N-H,H,N)}; 3:3 {(H,H,H-H,H,H), (H,H,H-H,HN), (H,H,N-H,H,N)}.

## 3. Results and discussion

### 3.1 Total scattering structure factors and partial radial distribution functions

Before discussing the results it is important to scrutinize (such as zeroth step) the 'goodness' of our structural models. Concerning the MD structure, we stress that the OPLS (six-site) force field already provided a very good, almost perfect fit to diffraction data, see Fig. 2: only slight differences can be detected on the neutron weighted TSSF. Thus this structural model proved to be a very good start for the subsequent RMC refinement.

We have to point out that the kind of comparison shown in Fig.2, simulation with measured total scattering structure factors, has either been missing completely from earlier simulation studies[7,9-13,15] or has shown only partial agreement[8,14] between simulation and experiment.

As it is obvious from Fig. 2, TSSF-s resulting from the RMC refinement are fully consistent with both the neutron- and X-ray diffraction data.

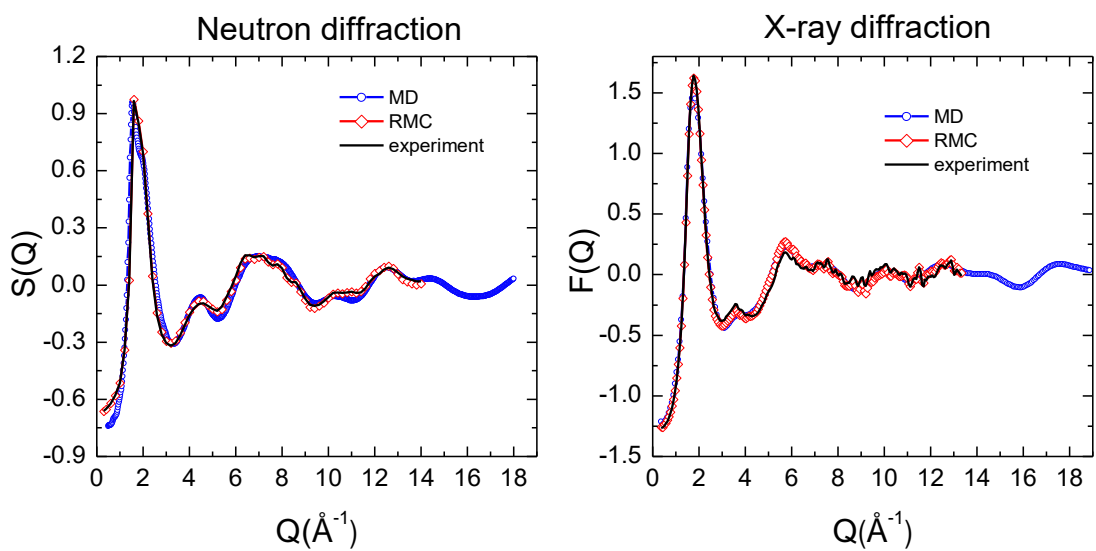

Figure 2. Total scattering structure factors for liquid $CH_3CN$. Blue line with empty circles: MD simulation; red line with empty diamonds: RMC model; black line: experiment.

Partial radial distribution functions calculated from the RMC models and the MD trajectories are compared in Fig. 3. As mentioned earlier, the two-particle distributions are represented by 10 partials as the two carbon atoms were distinguished. The MD based PRDF-s are reproduced well by RMC model: visible differences can be detected only in the C-N, H-N and H-H partials. It is reasonable to suggest that the slight differences between MD and RMC structures at the level of total scattering structure factors (cf. Fig. 2) mainly arise from these correlations.

First, concerning the six PRDF-s that do not contain hydrogen it can be stated that our findings are in very good agreement with previous molecular dynamics results of Böhm et al.[8] who also used a six-site potential model. Surprisingly, even Monte Carlo simulations of Jörgensen et al.[16] with a 3-site 'united atom' model provided concordant results. Although these PRDF-s have already been discussed in detail[8,16], an important observation may be highlighted here: the largest differences between different models[8,13,14,16,18] were observed in terms of the ratio of the intensities of the double peak of the C-N PRDF. Generally speaking, the first peak has a higher intensity for 3-site models[14,16,18], while in the case of six-site models[8,13] the second peak appears higher. In our case the MD simulation resulted in a higher first peak, while in the RMC model (in full agreement with diffraction data) the two peaks tend to show almost the same intensity.

Concerning pair correlations that involve hydrogen, the intra- and the intermolecular parts are well separated in the C-H, $C_{methyl}$-H and H-H PRDF-s, while this is not the case for the N-H partial (see Fig. 3). N-H intermolecular distances are smaller than the (non-bonded) intramolecular ones; this may be taken as an indication for antiparallel arrangements of neighbouring molecules. This feature has not been spotted in earlier studies.[8,13,14,16,18] We note that the sub-groups of the 'tetrahedral' orientational correlation functions (see Section 2.3.3, and below for related results) provide more information on the arrangements of hydrogen atoms than PRDF-s.

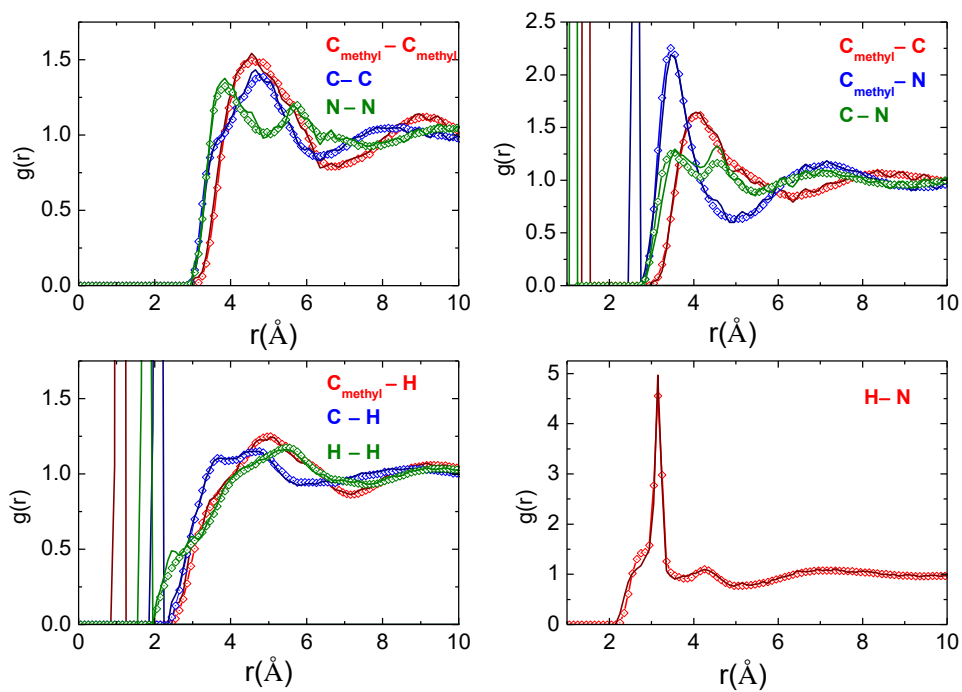

Figure 3. Partial radial distribution functions for liquid CH$_3$CN. Solid line: RMC; solid line with diamonds: MD results.

## 4.2 Dipole-dipole and 'tetrahedral' orientational correlations

In this subsection we aim to provide details of orientational correlations (introduced in Section 2.3) that can be found in the final RMC structural model of acetonitrile molecular liquid. The two families of correlation functions will be discussed in parallel, in order to emphasize essential features.

Dipole-dipole correlations are shown in Figure 4a. The strongest correlations, by far, appear for angles cca. 180 degrees; intensities significantly emerging from the background can also be seen around 90 and 0 degrees. Specific molecular axes correlations functions are therefore shown for three groups (Figs. 4b, 4c, 4d), according to their cos γ (cf. also Refs. 26, 27) value.

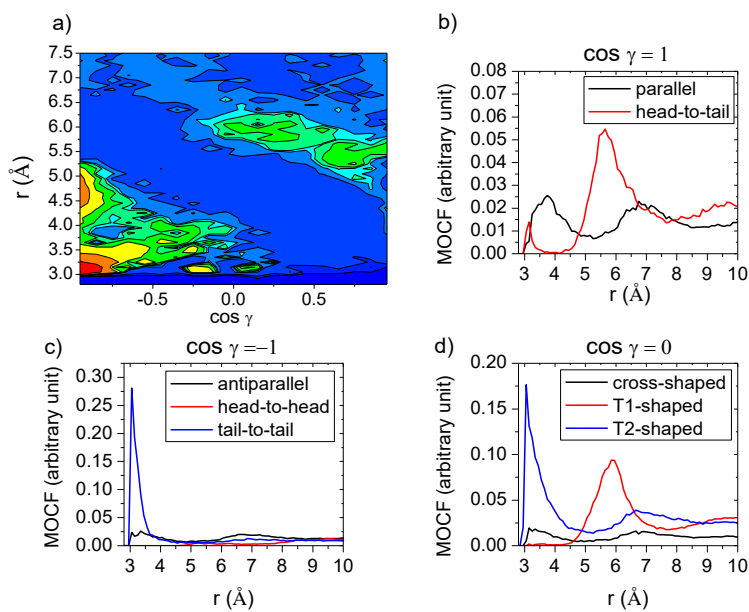

Figure 4. a) Dipole-dipole correlation functions. b) Molecular axes ('extended dipolar') correlation functions with cos γ =1. c) Molecular axes correlation functions with cos γ =−1. d) Molecular axes correlation functions with cos γ =0.

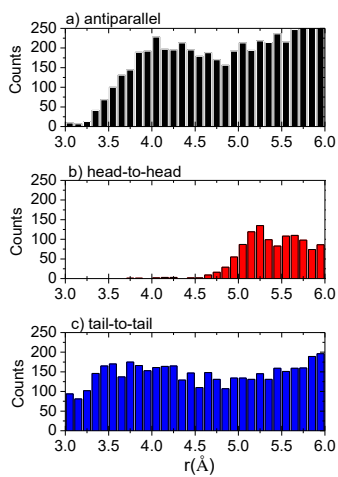

Figure 5. Un-normalized counts for the 'antiparallel' type arrangements. Top: Genuine, 'side-by-side' antiparallel; middle: head-to-head, i.e., for CH$_3$CN molecules, 'N-to-N'; bottom: tail-to-tail, i.e., 'methyl-to-methyl' configurations.

The strongest correlations between the dipolar vectors of two neighbour molecules show up below 3.5 Å, and are of the antiparallel type, where the cosine of the angle confined by the dipole axes is close to -1 (see Fig. 4a). There is another distinct spot in Fig. 4a that corresponds to antiparallel arrangements, between 4.5 and 5 Å. In order to find out the specific arrangements that belong to these maxima, one has to look at the corresponding specific dipole-dipole correlation functions, Fig. 4c, and the related un-normalized counts, Fig. 5. At short distances, it is clearly the occurrence of pairs of molecules that prefer to turn toward each other by their 'tail' (i.e., methyl-group) ends is most characteristic (cf. Figs. 4c and 4c): no other specific arrangement shows any intensity below 3.5 Å. In this case both cos α and cos β (for definitions of these angles, see Refs. 26, 27) are equal to zero. At the other important location, around 5 Å (cf. Fig. 4a), all the three antiparallel kinds of correlations, namely tail-to-tail (for acetonitrile, this means 'methyl-to-methyl'), head-to-head ('N-to-N') and the classic antiparallel 'side-by-side' ones contribute (see Fig. 5). We suggest that it is the sudden appearance of the head-to-head pairs (around 5 Å) that actually produces the intensity maximum in Fig 4a.

The above statements are in line with results on the 'tetrahedral' orientational correlation functions shown in Fig. 6. The 2:3 group has a dominant role below 3.5 Å and the probability of the 3:3 group also reaches 25% at the shortest centre-centre distances. It is worth pointing out that although it is the H,H,H-H,H,H subgroup of the 3:3 group that realizes the antiparallel case (cos γ =-1, cf. Fig. 4) most clearly, the H,H-H,H,H subgroup of the 2:3 group can also contribute to the 'antiparallel' area in Fig. 4a.

Another possibility to form a constellation in which methyl groups tend to be close to each other if cos γ, cos β are 0 and cos α is 1, see Fig. 4d: this is the so-called 'T2-shaped' arrangement (see Refs. 26,27). This constellation seems to be responsible for the small, but sharp maximum at short distances, in the vicinity of cos γ = 0 (Fig. 4a). Looking at the corresponding tetrahedral correlations (Fig. 6), the H,H-H,H,N group has a significant intensity, with almost 20% from the subgroups of the 2:3, and the H,H,H-H,H,N subgroup of the 3:3 group. That is, the importance of the proximity of the methyl groups is emphasized again at the shortest molecular centre – molecular centre distances (below cca. 3.5 Å).

Staying with the 'perpendicular' type arrangements (cos γ = 0), a noticeable maximum shows up around 6 Å (Fig. 4a): it can easily be spotted (Fig. 4d) that T1-shaped constellations are responsible for this feature (cos α=0, cos β=-1, cos γ=0; cf. Refs. 26,27))

Concerning parallel-like orientations, cos γ=1, the head-to-tail (cos α=1, cos β =-1, cos γ=1) orientation, with a well-defined maximum at 5.6 Å (see Fig. 4b), contributes significantly to the outstanding region of maximum intensities between 5 Å and 6.3 Å (Fig.

4a). Note that although the corresponding peaks (Figs. 4b and 4d) are spectacular, the characteristic intensity of this region is much lower than those which were found at the closest centre-centre distances.

In Fig.6 the oscillations of the orientational correlation functions rapidly decay, apart from the 1:2 (corner-to-edge) and 2:3 (edge-to-face) groups. These correlations show visible oscillations and further, they alternate far beyond 10 Å. Two remarks here: (1) these features (the oscillation and the alternation) can be found in the configurations already in the MD structure (without applying the RMC method); (2) similar characteristics were observed for $XY_4$ molecular liquids[42,43].

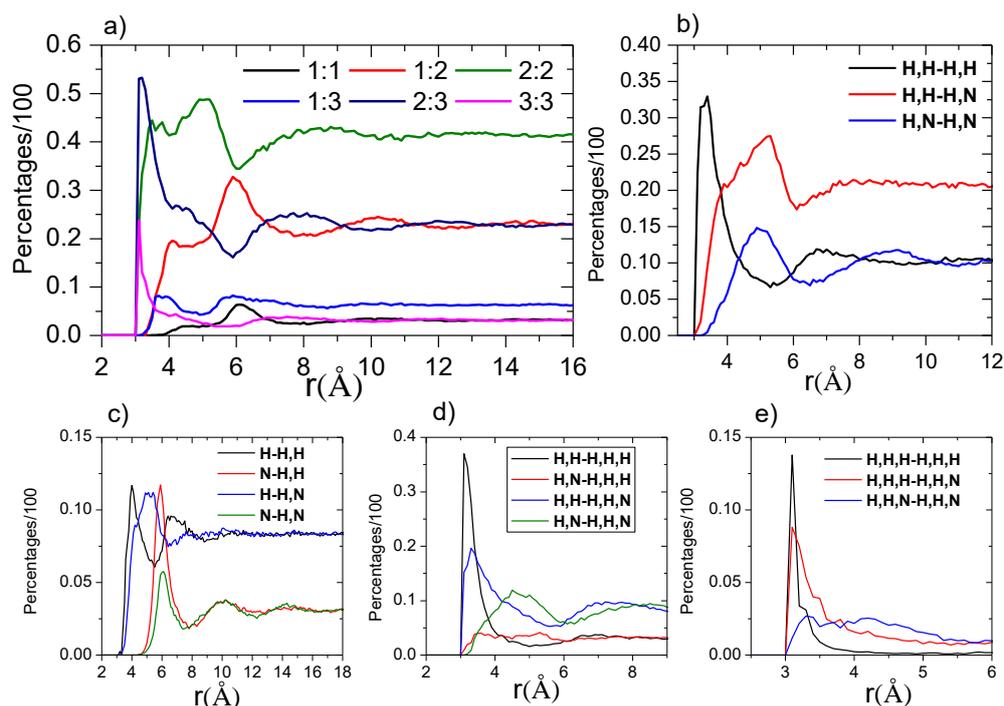

Figure 6: Orientational correlation functions. a) Six "original" groups. b) The subgroups of 2:2 (edge-to-edge) orientation. c) The subgroups of 1:2 (corner-to-edge) orientation. d) The subgroups of 2:3 (edge-to-face) orientation. e) The subgroups of 3:3 (face-to-face) orientation.

In general it is difficult to find an obvious connection between the positions of minima of any partial radial distribution functions and the positions of the minima of the dipole-dipole correlations functions (or the orientational correlation functions). An exception for liquid acetonitrile, and therefore worth noting, is the shoulder around 3.8 Å in the case of the C-C RDF (Fig. 3). (Note that this partial, with a very good approximation, can be considered as the centre-centre radial distribution function.) Around this shoulder we found a minimum in the dipole-dipole correlation functions (Fig. 4a), in the tail-to-tail function (Fig. 4c) as well as in the H,H,H-H,H,H (Fig. 6e) and the H,H-H,H,H (Fig. 6d) orientational correlation functions.

Concluding this section, it is worth noting that the mutual orientations of two neighbouring acetonitrile molecules were suggested to be antiparallel already in earlier studies[8,16,18,19,24], but beyond this distances range (above cca. 3.5 Å) a general consensus was missing [8,16,18,19,24]. Orientational correlation functions introduced above, together with special dipole-dipole correlation functions are found here a key tool to an accurate description how two acetonitrile molecules tend to orient relative to one another. It could be demonstrated, for example, that angular distributions or spatial distribution functions used previously were not able to make it possible to distinguish the head-to-tail arrangement from the 'side-by-side' parallel one, or the head-to-head, the tail-to-tail and 'side-by-side' antiparallel orientations from each other.[18]

As a final thought, we suggest that the kind of categorization of orientational correlations discussed above provides an opportunity for comparing liquid acetonitrile that is most frequently considered as a liquid with linear molecules, to liquids composed of tetrahedral molecules, such as carbon tetrabromide[43] or the members of the $XCl_4$ liquid family[44]. This is a novel approach in several aspects: just to mention one, usually the hydrogen atoms were united with the methyl carbon atom in earlier studies (e.g. 11, 14-20), thus it was difficult to gain information about correlation involving hydrogen. In the present study it was not our primary aim to provide a comprehensive comparison with tetrahedral liquids mentioned above – this may be the subject of a follow-up publication. To provide a fundamental link to tetrahedral systems, we mention here that the asymptotic values of the probabilities of the six main groups for $CH_3CN$ agree with those calculated for $XY_4$ liquids[41] to a very good approximation: the values are 3.2 (3.1 for $XY_4$) % (1:1), 22.95 (22.8) % (1:2), 41.5 (42.1) % (2:2), 6.2 (6.1) % (1:3), 22.95 (22.8) % (2:3), and 3.2 (3.1) % (3:3).

## 5. Conclusions

Based on Reverse Monte Carlo structural models that are fully consistent with both neutron and X-ray diffraction experimental data, orientations of molecules, especially the arrangements of hydrogen atoms in liquid acetonitrile were revealed.

As also stated in previous studies, antiparallel arrangements are the typical orientations of two neighbouring molecules. In this study, we were able to go beyond this, far too general, conjunction: special dipole-dipole correlations calculated here show that a significant number of these antiparallel molecular pairs are of the tail-to-tail type. Furthermore, in contrast to most of previous suggestions, T-shaped orientations are also found significant in the distance range up to about 3.5 Å. Increasing the distances between centres of molecules, first antiparallel orientations suddenly disappear (around 5.2 Å), the intensities of specific dipole correlation functions weaken, so that beyond the first coordination shell (cca. above 6.5 Å) any intensities became hardly detectable (see Fig 4a).

On the other hand, the tetrahedral approach introduced here also showed that the neighbouring molecules turn toward each other with their hydrogen sites. Concerning correlations beyond the first coordination shell, and also, the asymptotic values, they behave similarly as found previously for other liquids with molecules of tetrahedral shape, like carbon tetrachloride or chloroform.

Finally, it is worth highlighting once again that molecular dynamics simulations using the all-atom OPLS force field originally were meant to produce only initial configurations for RMC. During the process these simulation results, together with partial radial distribution functions obtained from them, have proven to be essential for Reverse Monte Carlo modeling and the subsequent analyses. Furthermore, for our MD model the quality of agreement with experimental total scattering structure factors exceeds the level showed for earlier MD simulations, and almost reaches that of the RMC simulation.

**Acknowledgements**
The authors are grateful to L. Temleitner for providing his unpublished X-ray diffraction data with us (X-ray beamtime was provided by the SPring-8 Facility at JASRI (Japan), under proposals No. 2011B1554 and 2014B1460). This work was funded by NKFIH (National Research, Development and Innovation Office of Hungary), Grant No. SNN 116198. Sz. P. wishes to acknowledge financial support of the Post Doctoral Research Programme of the Hungarian Academy of Sciences.